\documentclass[11pt, oneside, a4paper]{article}
\usepackage{graphicx}
\usepackage{amssymb}
\usepackage{url}
\usepackage[nocompress]{cite}
\usepackage{multirow}
\usepackage{array}
\usepackage{pavt}
\usepackage[labelsep=period]{caption}
\begin{document}

\title{A Distributed Parallel Algorithm for Minimum Spanning Tree Problem}

\authors{Artem Mazeev, Alexander Semenov, Alexey Simonov}

\organizations{JSC NICEVT, Moscow, Russia\\
\url{a.mazeev@nicevt.ru, semenov@nicevt.ru, simonov@nicevt.ru}
}

\begin{abstract}
In this paper we present and evaluate a parallel algorithm for solving a minimum spanning tree (MST) problem for supercomputers with distributed memory. The algorithm relies on the relaxation of the message processing order requirement for one specific message type compared to the original GHS (Gallager, Humblet, Spira) algorithm. Our algorithm adopts hashing and message compression optimization techniques as well. To the best of our knowledge, this is the first parallel implementation of the GHS algorithm that linearly scales to more than 32 nodes (256 cores) of Infiniband cluster.
\end{abstract}

\keywordsen{large graphs, MST, GHS, supercomputers, MPI}

\section{Introduction}
Given a connected, weighted undirected graph $G = (V, E)$, a spanning tree is a tree in this graph that contains all its vertices. Minimum Spanning Tree (MST)~\cite{mst} is a spanning tree having minimum possible weight, where the weight of the tree is the sum of the weights of all the edges contained in it.

The paper considers the minimum spanning tree problem in large graphs. By large graphs we mean graphs that can not fit in the memory of the typical node of the distributed memory system.

The MST problem is encountered in many areas, for example, in bioinformatics, computer vision and also when designing various networks. Requirements to the size of the processed graphs in real problems are constantly increasing. For example, in bioinformatics when solving clustering problem~\cite{lyubetsky} that can be solved by constructing a MST, graphs may take up to one petabyte or even more memory.

There are many algorithms~\cite{algs} that solve the MST problem; the best known algorithms are Prim's~\cite{prim}, Kruskal's~\cite{kruskal} and Boruvka's algorithms~\cite{boruvka}. Some algorithms are suitable for shared memory parallelization, there are lot of such implementations, for example~\cite{impl1, impl2, impl3, impl4}.

Some of the mentioned algorithms are adapted for implementation on distributed memory systems~\cite{pbgl, impl6, impl7, impl8, impl10}. Among the parallel implementations listed above there is not one implementation scalable to at least one hundred parallel processes. There are algorithms specially designed for distributed systems, for example, the GHS (Gallager, Humblet, Spira) algorithm~\cite{ghs} and Awerbuch~\cite{awerbuch}. To the best of our knowledge, there is only one paper~\cite{impl8} that describes the implementation of GHS algorithm, but it presents no good experimental results.

In this paper we present a parallel algorithm for solving a minimum spanning tree problem on distributed memory systems. The algorithm has been developed on the basis of the GHS algorithm. The algorithm allows processing of large-scale graphs and linearly scales to more than two hundred parallel processes.

\section{GHS Algorithm}
The GHS algorithm has been chosen for the study as a fundamental distributed parallel MST algorithm. This algorithm is based on a vertex-centric programming model~\cite{vertex_centric}. The idea of the algorithm is as follows: all vertices perform the same procedure, which consists of sending, receiving and processing the messages from adjacent vertices. The messages can be transmitted independently in both directions of an edge, the order of messages must be preserved along the edge direction.

At any time, the set of graph vertices is represented as a union of a certain number of fragments, i.e. the disjoint sets of vertices. Initially, each vertex is a fragment. Each fragment finds an edge with a minimum weight among the edges outgoing from this fragment to the other fragments. The fragments are then combined over these edges.
The edges, which are used to combine the fragments, will compose a minimum spanning tree when there is only one fragment comprising all the vertices left.

Consider the algorithm in detail.
There are three possible vertex states: \textit{Sleeping}, \textit{Find} and \textit{Found} where \textit{Sleeping} is the initial state of all vertices. The vertex will be in the state \textit{Find} when participating in a fragment's search for the minimum-weight outgoing edge, and in the state \textit{Found} in other cases. Each fragment has an $L$ variable characterizing its level. Initially the level of each fragment is 0. Two fragments of the same level $L$ can be combined into a level $L + 1$ fragment. A fragment cannot join to another fragment of a lower level.

The following is the detailed description of the searching process of the minimum-weight outgoing edge of the fragment. In the trivial case where the fragment consists of a single vertex and its level is 0, the vertex locally chooses its minimum-weight outgoing edge, marks this edge as a branch of the minimum spanning tree and sends a message called \textit{Connect} over this edge and goes into the \textit{Found} state.

Now consider the case where a fragment level is greater than 0. Suppose a new fragment at level $L$ has just been formed by the combination of two level $L - 1$ fragments with the same outgoing edge, which becomes the core of the new fragment. The weight of this core edge is used as the identity of the fragment. Then an \textit{Initiate} message is broadcast all over the fragment starting from the vertices adjacent to the core, so that all vertices receive new fragment level and identity and are placed in the \textit{Find} state. When a vertex receives the \textit{Initiate} message, it starts finding the minimum-weight outgoing edge.

Each edge of the graph can be in one of three states: \textit{Branch}, if the edge belongs to the minimum spanning tree; \textit{Rejected}, if the edge is not part of the mininum spanning tree; and \textit{Basic} if it is not yet known whether the edge is part of the minimum spanning tree or not. In order to find minimum-weight outgoing edge, for each vertex $v$ all edges in the \textit{Basic} state are sorted out starting from the most light-weight edge. Each edge is probed by sending \textit{Test} messages along that edge.
The \textit{Test} message contains fragment level and identity as arguments. When vertex $u$ receives the \textit{Test} message, it compares its own fragment identity with one received in the message. If the identities are equal, then the vertex $u$ sends the \textit{Reject} message back, and then both vertices put the edge in the \textit{Reject} state. In this case, the vertex $v$ that has sent the \textit{Test} message, continues the search, analyzing the next best edge and so on. If the fragment identity in the received \textit{Test} message is different from the fragment identity of the receiving vertex $u$, and if the receiving vertex fragment level is greater or equal to the one in the \textit{Test} message, then an \textit{Accept} message is sent back. In this case the state of the edge incident to the vertex $v$ is changed to \textit{Branch}. However, if the fragment level of the vertex $u$ is smaller than the one in the received message, then the message is postponed, until the fragment level of the vertex $u$ increases to the necessary value.

\begin{figure}[h]
\begin{minipage}[h]{0.49\linewidth}
\center{\includegraphics[width=1\linewidth]{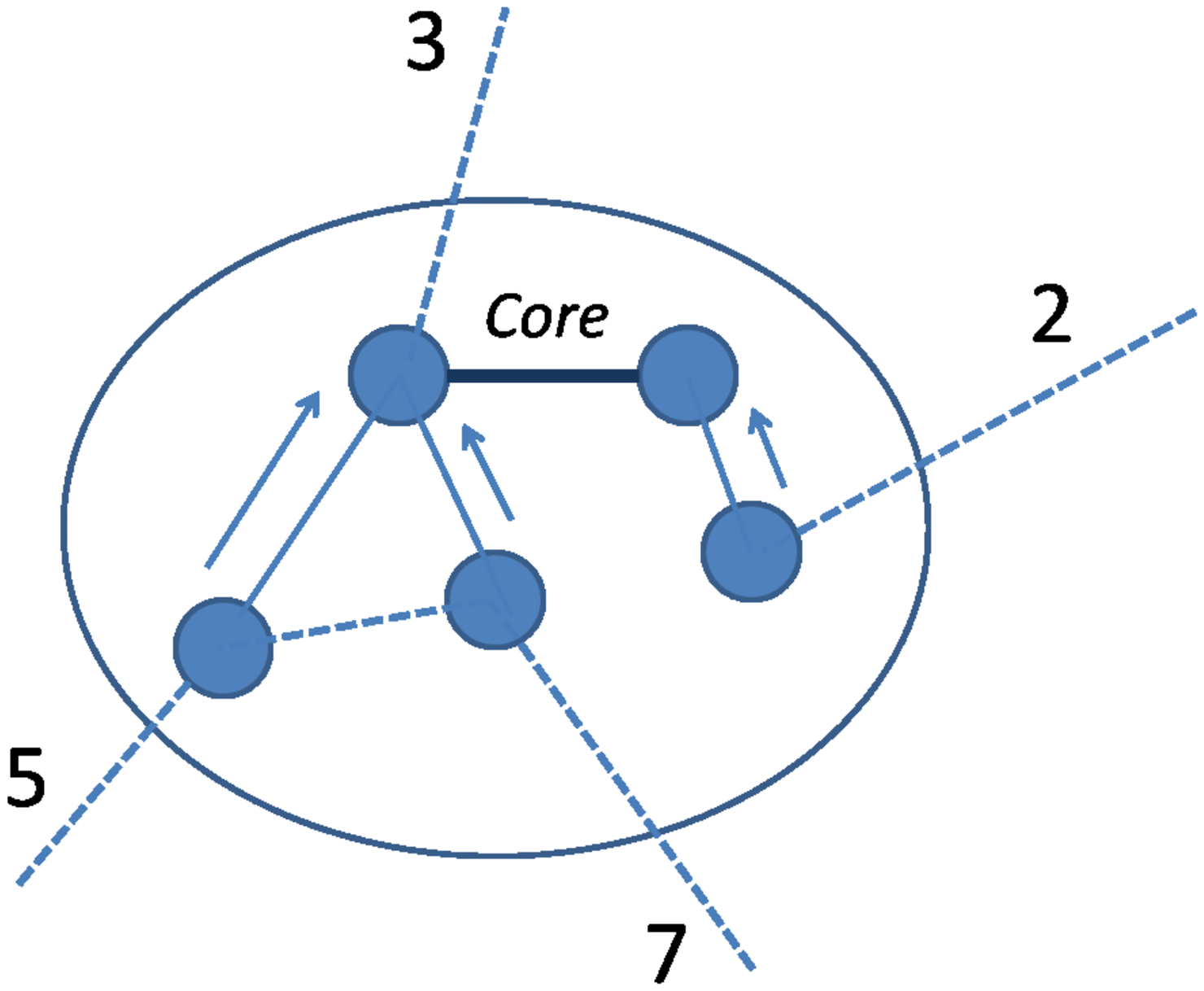} \\ a)}
\end{minipage}
\hfill
\begin{minipage}[h]{0.49\linewidth}
\center{\includegraphics[width=1\linewidth]{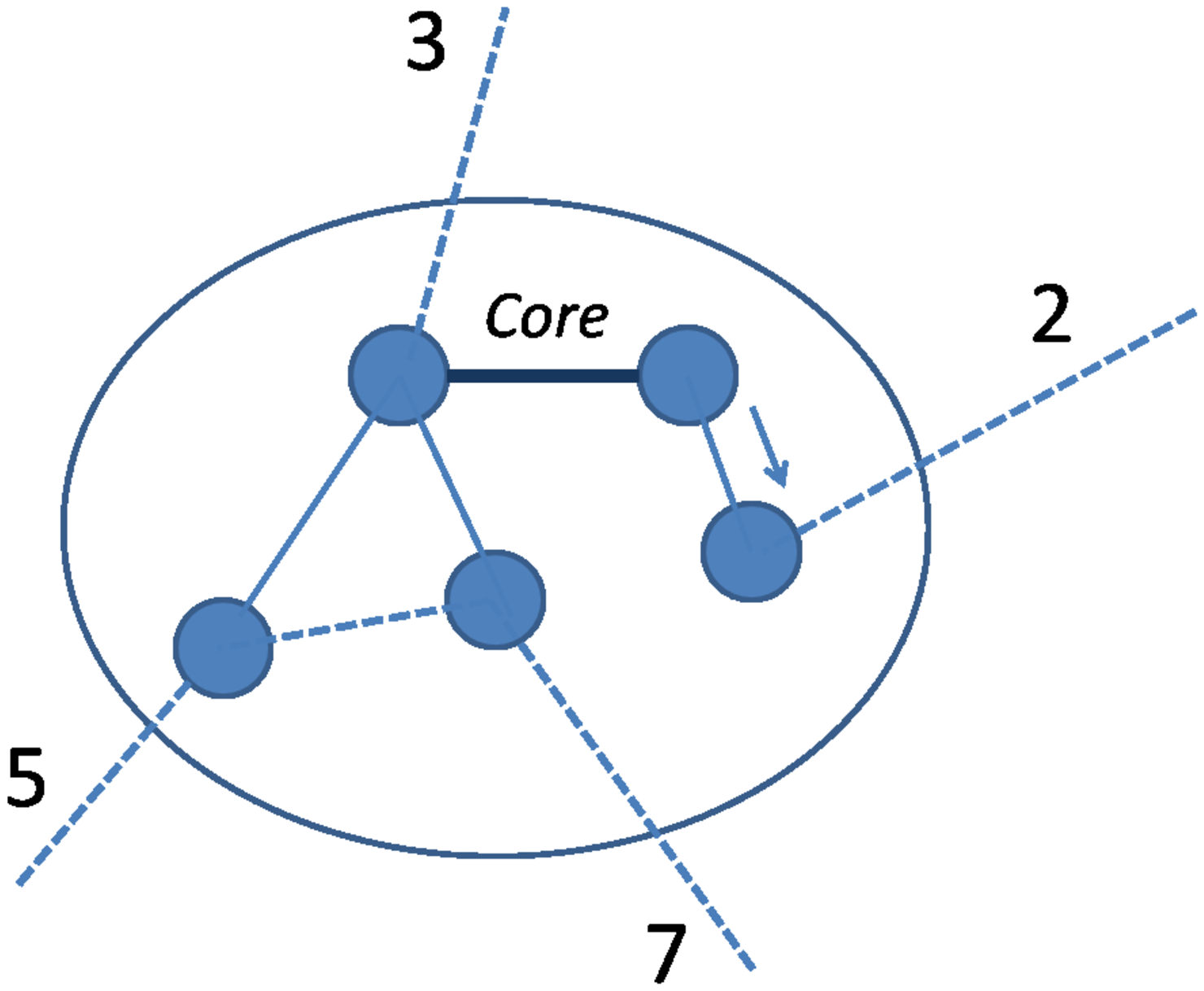} \\ b)}
\end{minipage}
\caption{The scheme of the GHS algorithm execution. In Fig. a) arrows denote the sending of the \textit{Report} messages. In Fig. b) arrow denotes the sending of the \textit{Change core} message towards the minimum-weight outgoing edge of the fragment. The edges shown by solid lines are in the \textit{Branch} state, numbers on edges are weight values.}
\label{ris:ghs}
\end{figure}

Finally each vertex finds a minimum-weight outgoing edge, if any. Now vertices are sending \textit{Report} messages (see.~Fig.~\ref{ris:ghs}.a)~), to find the minimum-weight outgoing edge of the whole fragment. If none of the graph vertices have outgoing edges in the \textit{Basic} state, then the algorithm terminates, and the edges in the \textit{Branch} state are the minimum spanning tree.

\textit{Report} messages are sent by the following rules. Each leaf vertex of the fragment sends \textit{Report(w)} along the only incident edge in the \textit{Branch} state (\textit{w} is a weight of the minimum outgoing edge from the vertex or infinity, if there are no outgoing edges). Each internal vertex finds its own minimum-weight outgoing edge and waits for all messages from all the subtrees. Then the vertex chooses minimum weight from all the weight values. If the minimum is achieved with the value which came from the subtrees, then a number of the outgoing branch is put into vertex variable \textit{best\_edge}, otherwise a number of the minimum-weight outgoing edge is put into this variable. This is done in order to easily restore the path by moving to where the \textit{best\_edge} is pointing. Further there is a sending of the \textit{Report} message up the tree of the fragment with an argument equal to the found minimum value among all the weight values. When the vertex sends the \textit{Report} message, it also goes into the \textit{Found} state. Finally, two vertices that are incident to the core edge send the \textit{Report} messages along the core and determine the weight of the minimum outgoing edge and the direction to this edge.

In order to try to connect one fragment to another over the found minimum-weight outgoing edge of the fragment, it is possible to use \textit{best\_edge} variable in every vertex to trace the path from the core to the minimum-weight outgoing edge. For this purpose, a \textit{Change core} message is sent from one of the core vertices that is closer to the minimum-weight outgoing edge (see.~Fig.~\ref{ris:ghs}.b)~). A vertex that has received this message, sends it further in accordance with its own \textit{best\_edge} value, and so on. When the message reaches the vertex having minimum-weight outgoing edge, then this vertex becomes the root of the tree formed by the fragment. This vertex sends the \textit{Connect(L)} message over the minimum-weight outgoing edge, where $L$ is a fragment level. If two level $L$ fragments have the same minimum-weight outgoing edge, then each of them sends the \textit{Connect(L)} message over this edge, and this edge becomes the core of the new level $L + 1$ fragment, which immediately starts to send the \textit{Initiate} message with a new level number and identity all over the fragment.

When a level $L$ fragment with an identity $F$ sends the \textit{Connect} message into the level $L' > L$ fragment with identity $F'$, the larger fragment will send the \textit{Initiate} message with $L'$ and $F'$ into the smaller fragment.

Time complexity of the GHS algorithm is $O(N\log N)$, the number of communication messages is $5N\log N+2M$, where $N$ is a number of vertices, $M$ is a number of edges in the graph.
Not all the occurring cases are considered in this algorithm description but only the basic ones.

\section{MST Algorithm}
GHS algorithm presented in the paper of 1983~\cite{ghs} is only a description and analysis of the necessary high-level steps, that must be performed at every vertex. 
As far as we know, there is no paper that describes implementation details of the algorithm and scales well. 

It is necessary to reasonably choose and develop a set of techniques and to solve a number of problems for the development of parallel algorithm for finding an MST based on GHS algorithm. Implementation of the proposed algorithm has been made using C++ language with an MPI library. When running on a supercomputer the number of vertices in the graph is significantly larger than the number of MPI-processes, so a large number of vertices and all related information are typically stored in the memory of each process. All graph vertices are sequentially distributed in blocks among the processes. The local part of the graph in each process is stored in the CRS (Compressed Row Storage) format.

\subsection{Preprocessing of the Original Graph}
Preprocessing of the graph is conducted before searching for minimum spanning tree in the graph: loops and multiple edges are removed from the graph. The removal of multiple edges is used to fulfill GHS algorithm condition which says that all the edges must be unique. The time spent on the preprocessing is negligible and not included in the total time of algorithm execution. 

\subsection{Base Version}
\label{lab:base_version}

The base version of the algorithm has been developed at the beginning of work. Every MPI-process supports a queue where vertices can postpone a message if it is necessary. The aggregation of messages is implemented to speed up the algorithm; a separate buffer is created in every process for every possible receiving process.

\medskip

\noindent
\textit{Implementation scheme of the base version of the parallel algorithm for construction of an MST using MPI library; executed in parallel at every MPI-process.}
\begin{verbatim}
Input: local_G - local part of the graph
Output: local part of the MST

While (True) {
    /* read messages and push them to the queue */
    read_msgs ();
    /* queue processing, sending messages (write to the send buffer) */
    If (time_to_process_queue) {
        process_queue ();
    }
    If (time_to_send) {
        /* send all aggregated messages */
        send_all_bufs ();
    }
    /* checking for algorithm completion using MPI_Allreduce */
    check_finish ();
}
\end{verbatim}

Besides information that is necessary for algorithm execution messages also contain service information: the number of sending vertex and the number of the receiving vertex, as well as the message type. 

It is important to note that the GHS algorithm requires original graph to be connected. It is not necessary for the proposed algorithm because it will work until the interconnect is in the "silence" state, when all queues are empty, all messages are processed and there are no undelivered messages in the network.
Thus, the proposed algorithm allows finding not only an MST in a connected graph, but also a minimum spanning forest in the graph with any number of connected components.

Since GHS algorithm requires the weights of graph edges to be different, a special identity $special\_id$ is added to the usual weight of the edge. $special\_id$ for every graph edge $e$ is calculated as follows: let $u$ and $v$ be the vertices that are incident to the edge $e$, then $special\_id$ in binary representation equals to the consecutively recorded binary representations of $min(u, v)$, $max(u, v)$. Such an arrangement enables algorithm to work correctly even if the input graph has two different edges with the same weights.

\subsection{Searching Local Edges}
\label{lab:find_version}
When MPI process received an incoming message, it is necessary to find the edge (an edge index in the list of local edges) over which the message came, i.e. to find an index of the edge formed by the two vertices (sending and receiving) in the list of local edges. The search is necessary because the change of the local data related to that edge may be required.

The base version uses a linear search for this operation. During linear search all edges that are incident to receiving vertex are sorted out. If the vertex on the other end of the edge is equal to the sending vertex, then the right edge is found.

The first possible way to optimize this operation was sorting of all incident edges at every vertex of the original graph in increasing order of vertices numbers on the opposite end of the edge. With such an approach, at the beginning of the algorithm execution it is necessary to spend a little time on sorting, but during the algorithm execution a binary search can be used instead of linear. Such an approach provides a small gain in performance.

The second optimization that was considered is hashing. It is possible to create a hash table in every process instead of sorting and binary search. Let $u$ be the vertex sending a message, $v$ is receiving vertex, vertex identifier is a 32 bit machine word. Let’s define a hash function $get\_hash(u, v)$ as
\begin{equation}
((u \ll 32)\; | \; v) \; \textbf{mod} \; hash\_table\_size,
\end{equation}
where $\ll$ is a bit left shift, $|$ is a bitwise OR, $\textbf{mod}$ is a remainder of division, $hash\_table\_size$ is a hash table size (several times larger than the number of local edges).

Hashing method used in the proposed algorithm is called \textit{linear search and insertion}~\cite{hashing}. Thus an identity of the local edge can be found on two adjacent vertices for within $O(1)$, but first it is necessary to create and populate the hash table. This procedure is a part of the algorithm initialization and takes very little time and is not included in the total time of algorithm execution.

\subsection{Separate Processing of the \textit{Test} Messages}
\label{lab:test_version}
It is not always possible to immediately process certain types of messages (\textit{Connect}, \textit{Test} and \textit{Report}), because several conditions must be satisfied to perform the processing. The condition would be satisfied when some data changes, and to change the specific data it is necessary to wait for a specific message. So, there are situations when a message should be postponed, and then an attempt to process it again should be made. It is not known when it will be processed. 

Original GHS algorithm requires the preservation of the messages order, but the study of the algorithm execution showed that \textit{Test} messages constitute a significant part of all messages. It was found that it is beneficial to organize a separate queue for \textit{Test} messages, and to process it much less frequently than the main queue.

\subsection{Messages Length Optimization}
\label{lab:msg_opt_version}
To achieve maximum possible performance of the algorithm implementation on a distributed memory system it is necessary to minimize the size of the communication messages. It is therefore important that the structure that stores the message takes as little memory as possible.

At first messages were grouped into "short" (\textit{Connect}, \textit{Accept}, \textit{Reject}, \textit{Change core}) and "long" messages (\textit{Initiate}, \textit{Test}, \textit{Report}). The main difference is that "long" messages contain the weight and it takes significant amount of memory (64 bit).

In the beginning of each structure, for both "long" and "short" messages, a packed bit field of 16 bits is stored (actually only 9 bits are necessary: 3 bits for message type, 5 bits for fragment level, 1 bit for vertex state). Further, the structure stores the identifiers of sending vertex and receiving vertex (vertex identifier is a 32 bit machine word). Long messages further store the extended weight of the edge ($special\_id$) and the weight itself.

Finally the following optimization is implemented. Instead of storing $special\_id$ in the extended weight  (concatenation of two vertices identifiers, 64 bits in total), it is possible to store the minimal number from all the numbers of MPI processes which store this edge after verifying that the weights of all the edges in every process are different. Indeed, if the weights of all edges in every process are different, then two different edges with the same weights can only be in different processes, but then, the numbers of relevant processes are enough to understand that such edges are different.

As a result short and long messages are 80 and 152 bits size respectively.

\subsection{Parameters of the Proposed Algorithm}
\label{lab:params}
There are relevant implementation parameters:
\begin{itemize}
\item $MAX\_MSG\_SIZE$ is the maximum size of aggregated messages (by default, 10000 bytes),
\item $SENDING\_FREQUENCY$ is the frequency of flushing aggregated messages (by default, every 5 iterations of $while$ loop),
\item $CHECK\_FREQUENCY$ is the frequency of processing the queue with $Test$ messages (by default, every 5 iterations of $while$ loop), 
\item $EMPTY\_ITER\_CNT\_TO\_BREAK$ is the frequency of checking for completion (by default, every 100000 iterations of $while$ loop),
\item $HASH\_TABLE\_SIZE$ is the size of the hash table, in number of elements. Default value is $local\_actual\_m~*~5~*~11~/~13$, where $local\_actual\_m$ is the number of local edges in the MPI process after removing multiple edges and self-loops.

\end{itemize}

\section{Experimental Results}

RMAT, SSCA2 and Uniformly Random graphs are used for performance evaluation of the algorithm.
\begin{itemize}
\item RMAT~\cite{rmat} graphs represent real-world large-scale graphs from social networks and Internet, and are complex enough to analyze, so they are often used to evaluate performance of graph processing algorithms.
\item SSCA2~\cite{ssca2} graphs represent set of randomly connected cliques.
\item In Uniformly Random~\cite{random-uniform} graphs neighbours of each vertex are chosen randomly.
\end{itemize} 

The paper examines graphs with an average vertex degree of 32 and a pow of 2 number of vertices. Weights of the edges are a real numbers in the (0, 1) interval. $SCALE$ parameter specifies the number of vertices in the graph. If $n$ is the $SCALE$ parameter, then $2^n$ is the number of vertices in the graph. Graph with $SCALE=n$ is hereinafter referred to as, for example, RMAT-$n$.

The default values of the algorithm parameters listed in subsection \ref{lab:params} are used for performance evaluation.

We focus our design and experimental evaluation on the MVS-10P cluster system. Table~\ref{tab:systems} provides an architecture overview of the system.

\begin{table}[h]
\caption{\label{tab:systems}MVS-10P cluster system configuration.}
\begin{center}
\renewcommand{\arraystretch}{1}
\begin{tabular}{|p{33mm}|p{60mm}|}
\hline
 & \textbf{MVS-10P}\\
\hline
\textbf{Nodes} & 2x Xeon E5-2690 (8 cores, 2.9 GHz)\\
\hline
\textbf{Number of nodes} & 207\\
\hline
\textbf{Memory} & 64 GB\\
\hline
\textbf{Interconnect} & Infiniband 4xFDR\\
\hline
\textbf{MPI} & Intel MPI 4.1\\
\hline
\end{tabular}
\end{center}
\end{table}

\subsection{Impact of Optimizations}
In this subsection RMAT graph with scale 23 was used for testing (RMAT-23). The number of MPI processes per one node of the MVS-10P cluster is 8.

If binary search is used instead of linear search when finding a local edge, then the execution time on the cluster node is reduced by 2\%, if hashing is used instead of linear search, then the execution time on the node is approximately 18\% less (MVS-10P cluster, RMAT-23 graph, 8 MPI processes per node). Thus, the option of hashing was chosen for the final version.

\begin{figure}[h]
\begin{minipage}[h]{0.49\linewidth}
\center{\includegraphics[width=1\linewidth]{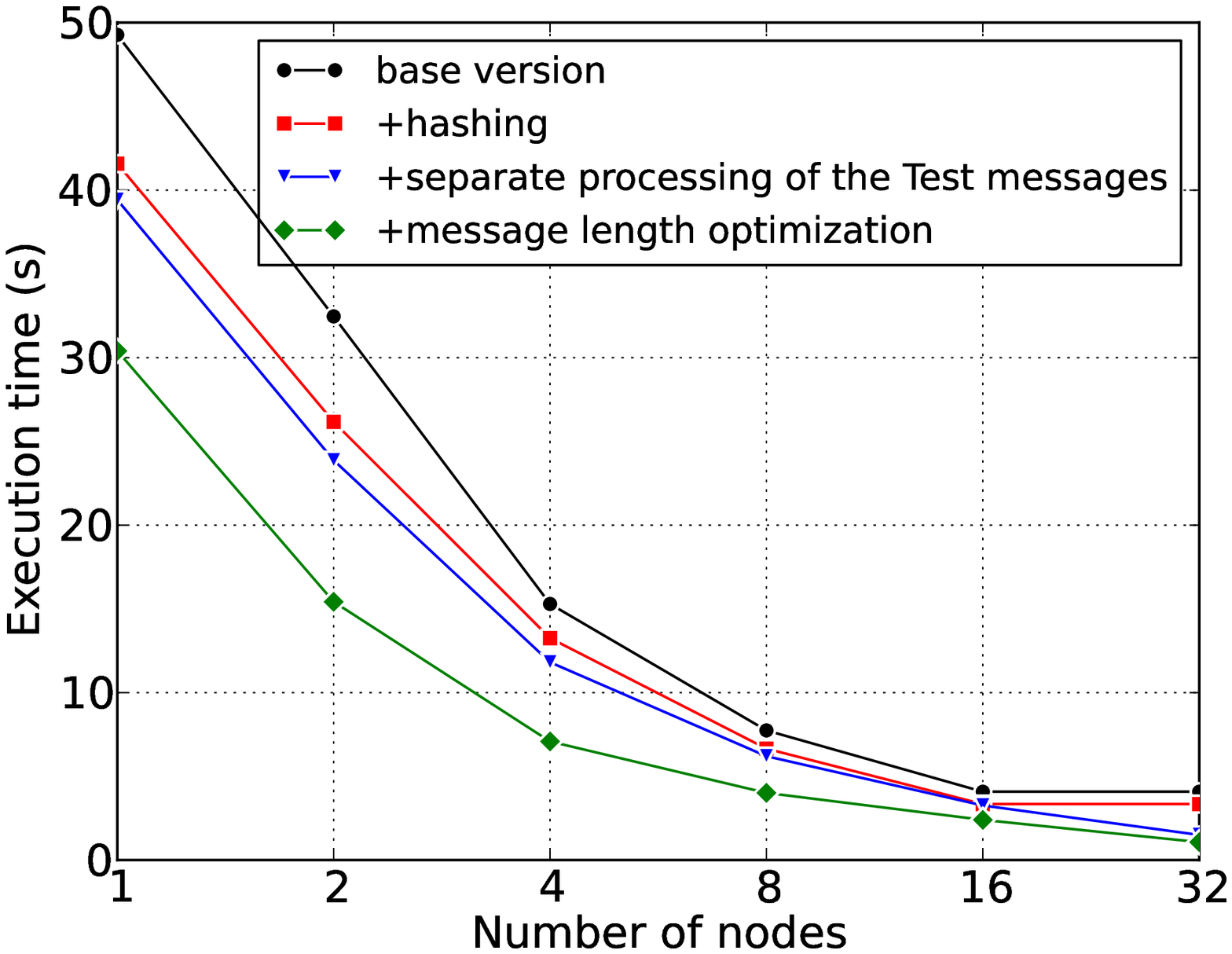} \\ a)}
\end{minipage}
\hfill
\begin{minipage}[h]{0.49\linewidth}
\center{\includegraphics[width=1\linewidth]{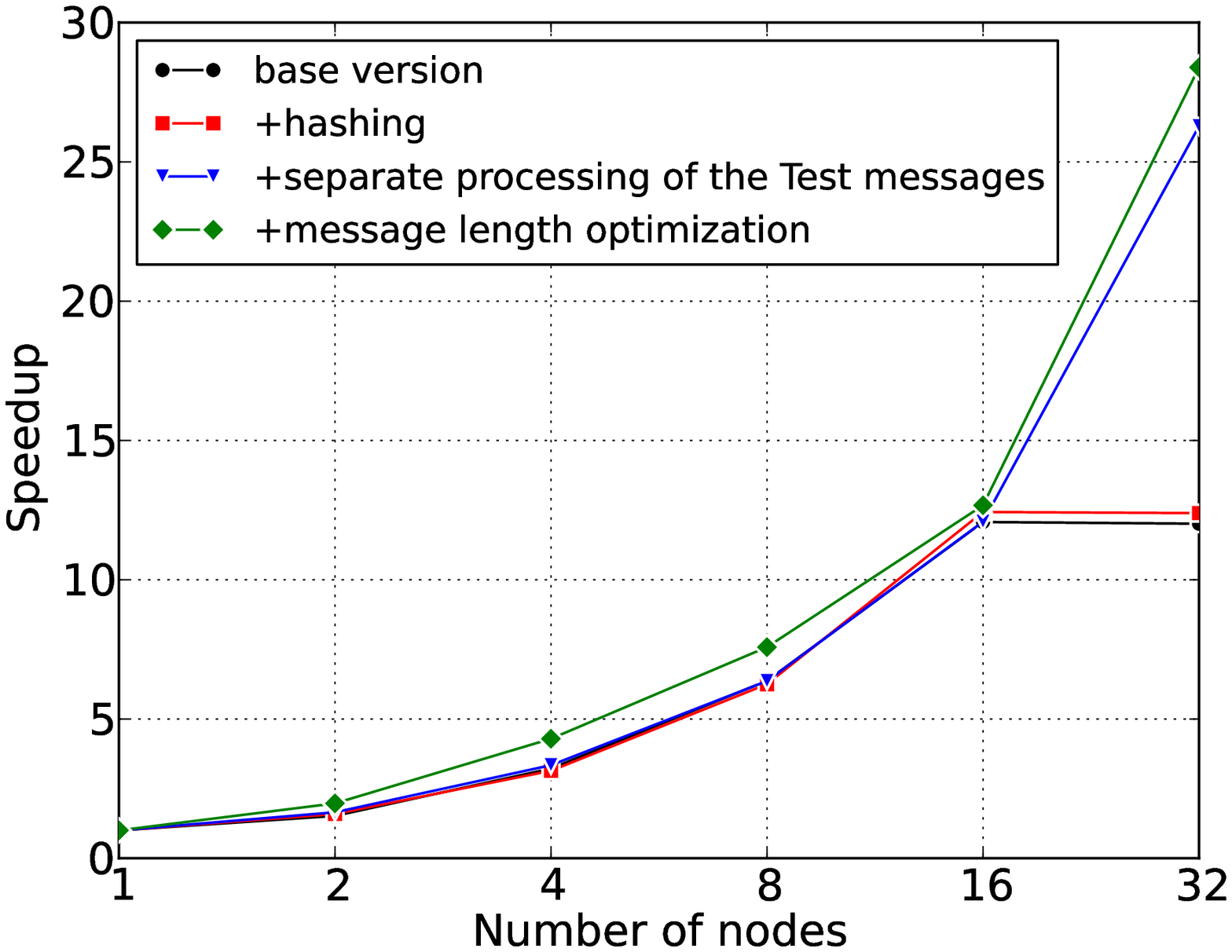} \\ b)}
\end{minipage}
\caption{Impact of optimizations: from the base version to the final version (with all the optimizations). MVS-10P cluster, RMAT-23 graph, 8 MPI processes per node.}
\label{ris:rmat23_all}
\end{figure}

Fig.~\ref{ris:rmat23_all}.a) shows how the runtime has been changing (in seconds), as the optimizations described in \ref{lab:find_version}, \ref{lab:test_version}, \ref{lab:msg_opt_version} have been added.
~Fig.~\ref{ris:rmat23_all}.b) shows the scalability of the same runs, i.e., the ratio of the problem solution time on one node to the problem solution time on a given number of nodes.

\begin{figure}[h]
\begin{minipage}[h]{0.49\linewidth}
\center{\includegraphics[width=1\linewidth]{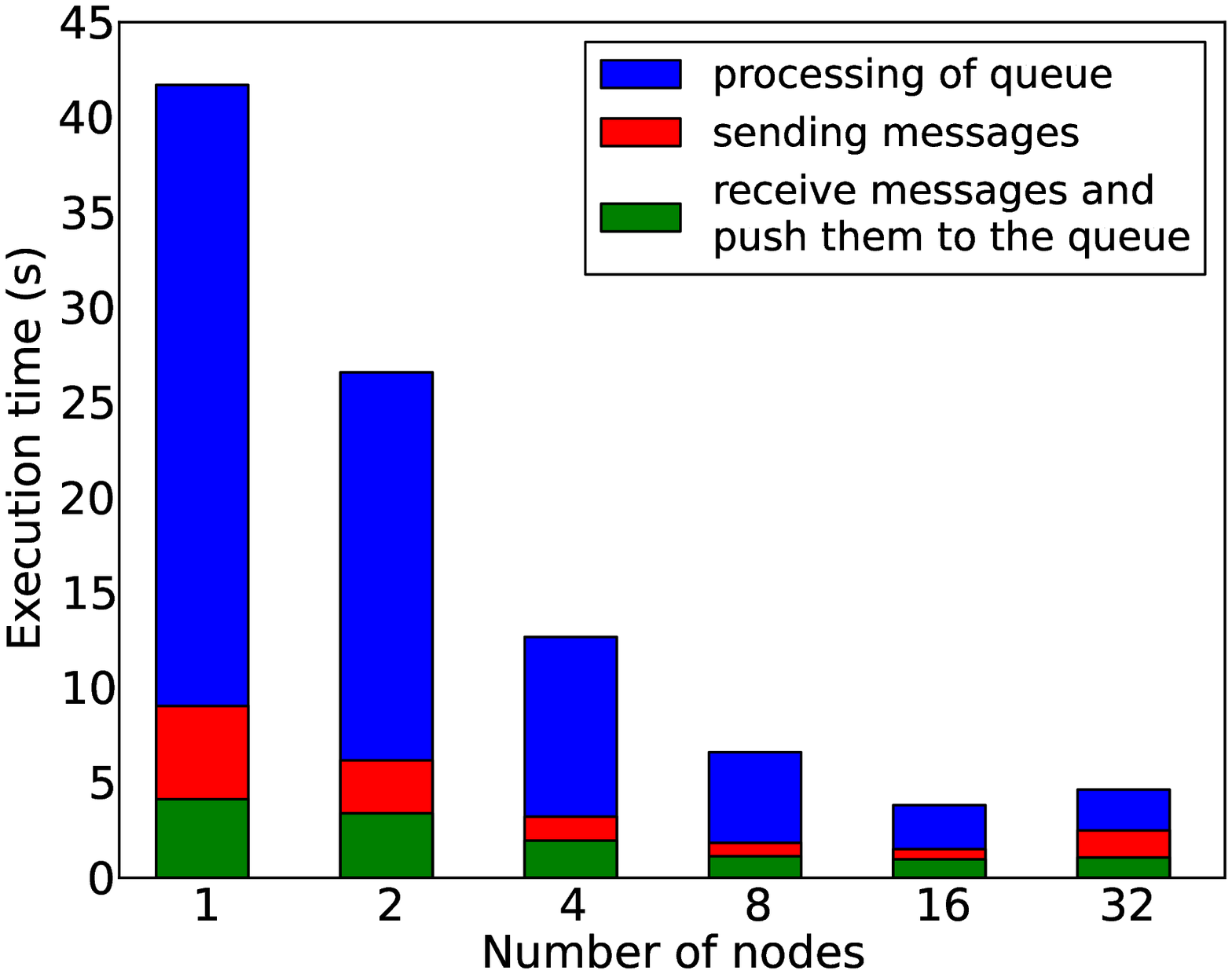} \\ a) Version with hashing}
\end{minipage}
\hfill
\begin{minipage}[h]{0.49\linewidth}
\center{\includegraphics[width=1\linewidth]{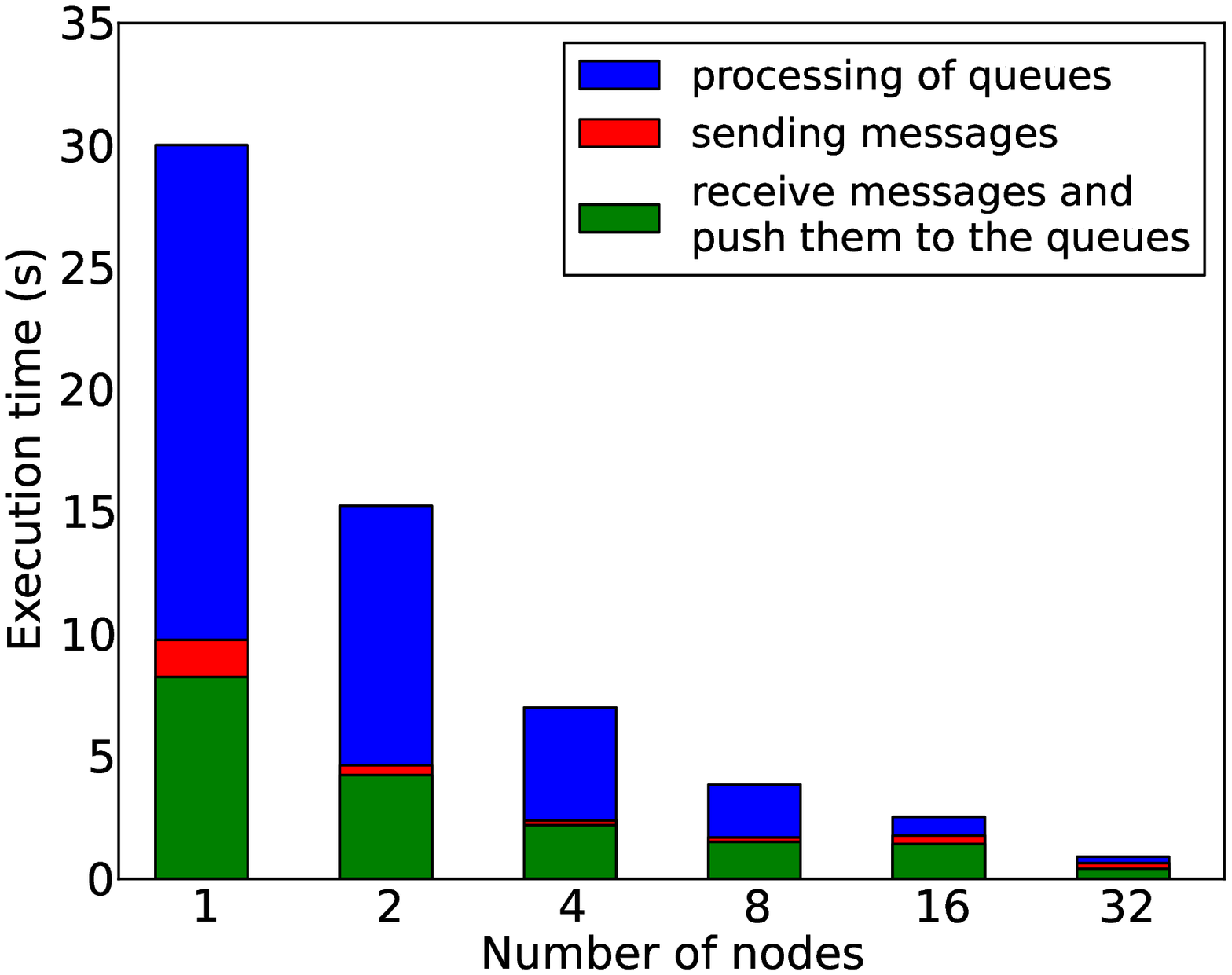} \\ b) Final version}
\end{minipage}
\caption{Profiling results. MVS-10P cluster, RMAT-23 graph, 8 MPI processes per node.}
\label{ris:timing_profile}
\end{figure}

Fig.~\ref{ris:timing_profile}.a)\hspace{1mm} shows the profiling results of the algorithm version with one optimization of the local edge search, and~Fig.~\ref{ris:timing_profile}.b)\hspace{1mm} shows the profiling results of the final version of the algorithm.

The profiling shows that the most of the time is spent on processing of queues. Some messages are processed repeatedly including \textit{Test} messages, so in the final version of the algorithm in which \textit{Test} messages are processed less frequently the part of queue processing in total execution time of the algorithm is less than in version with only hashing optimization. Exactly this optimization improved the algorithm scalability by 2 times, see~Fig.~\ref{ris:rmat23_all}.b).

Also, message length optimization has made a considerable contribution to performance of algorithm implementation. This optimization reduced the execution time of the final version of the algorithm on any number of nodes by approximately 50\%.

\subsection{Scaling}
Table~\ref{tab:perf_vert} shows performance evaluation results of the final algorithm version on the MVS-10P cluster for RMAT-24, SSCA2-24 and Random-24 graphs. Number of MPI processes per node is 8.

\begin{table}[h!]
\caption{\label{tab:perf_vert}The performance of the proposed algorithm on the MVS-10P cluster system. The scale of all used graphs is 24.}
\begin{center}
\renewcommand{\arraystretch}{1}
\begin{tabular}{|>{\centering}p{19mm}|>{\centering}p{21mm}|>{\centering}p{17mm}|>{\centering}p{8mm}|>{\centering}p{8mm}|>{\centering}p{8mm}|>{\centering}p{8mm}|>{\centering}p{8mm}|>{\centering}p{8mm}|>{\centering\arraybackslash}p{8mm}|}
\hline
\multicolumn{3}{|c|}{\textbf{Number of nodes}} & \textbf{1} & \textbf{2} & \textbf{4} & \textbf{8} & \textbf{16} & \textbf{32} & \textbf{64}\\
\hline

\multirow{6}{*}{\textbf{MVS-10P}} & \multirow{2}{*}{\textbf{RMAT-24}} & \textbf{Time {(s)}} & 63,27 & 36,12 & 17,98 & 8,47 & 5,41 & 2,04 & 1,45 \\
\cline{3-10}
& & \textbf{Scaling} & 1,00 & 1,75 & 3,52 & 7,47 & 11,7 & 31,01 & 43,63 \\
\cline{2-10}
& \multirow{2}{*}{\textbf{SSCA2-24}} & \textbf{Time {(s)}} & 54,69 & 32,37 & 11,90 & 6,02 & 3,63 & 1,72 & n/a \\
\cline{3-10}
& & \textbf{Scaling} & 1,00 & 1,69 & 4,60 & 9,08 & 15,07 & 31,62 & n/a \\
\cline{2-10}
& \multirow{2}{*}{\textbf{Random-24}} & \textbf{Time {(s)}} & 88,61 & 51,65 & 21,47 & 10,27 & 6,68 & 3,23 & n/a \\
\cline{3-10}
& & \textbf{Scaling} & 1,00 & 1,72 & 4,13 & 8,63 & 13,26 & 27,43 & n/a \\
\hline


\end{tabular}
\end{center}
\end{table}

$SCALE$ 24 is the largest graph scale that fits into the memory the MVS-10P node. The size of these graphs is approximately 6.5 GB. The rest of the memory node is needed for algorithm implementation. In particular, a large amount of memory is required to organize the hash table. 

Scalable mode in Intel MPI 4.1 on the MVS-10P cluster provides linear scaling on 32 nodes. On 64 nodes (512 cores) of the MVS-10P the scaling is 43.6. 

\begin{figure}[h]
\begin{center}
\begin{minipage}[h]{0.48\linewidth}
\includegraphics[width=1\linewidth, height=5cm]{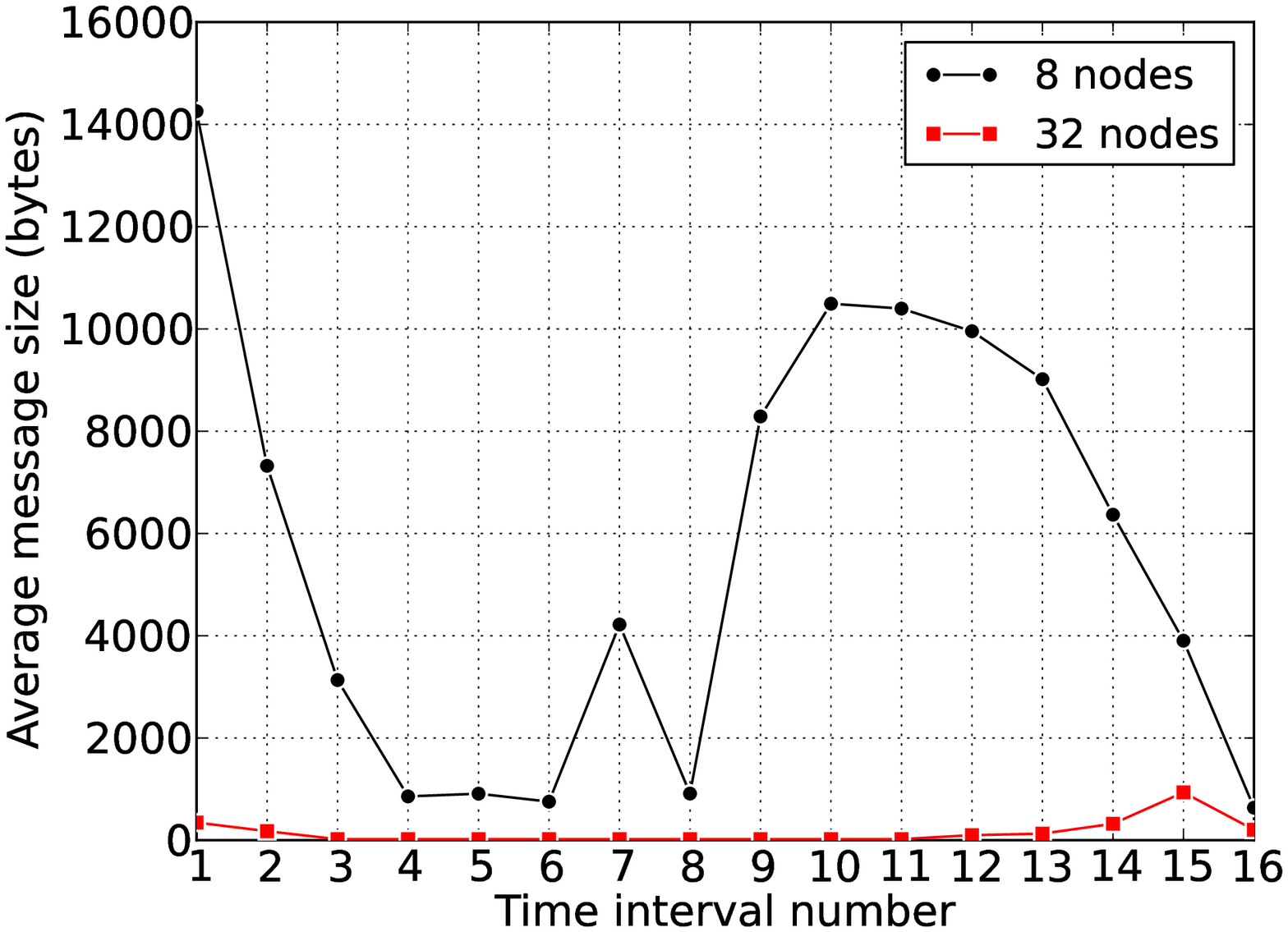}
\caption{The average size (over all MPI processes) of communication messages in bytes depending on the interval number (total execution time of the algorithm is divided into equal intervals). MVS-10P cluster, RMAT-23 graph, 8 MPI processes per node.}
\label{ris:avg_msg_size}
\end{minipage}
\hfill 
\begin{minipage}[h]{0.48\linewidth}
\includegraphics[width=1\linewidth,height=5.5cm]{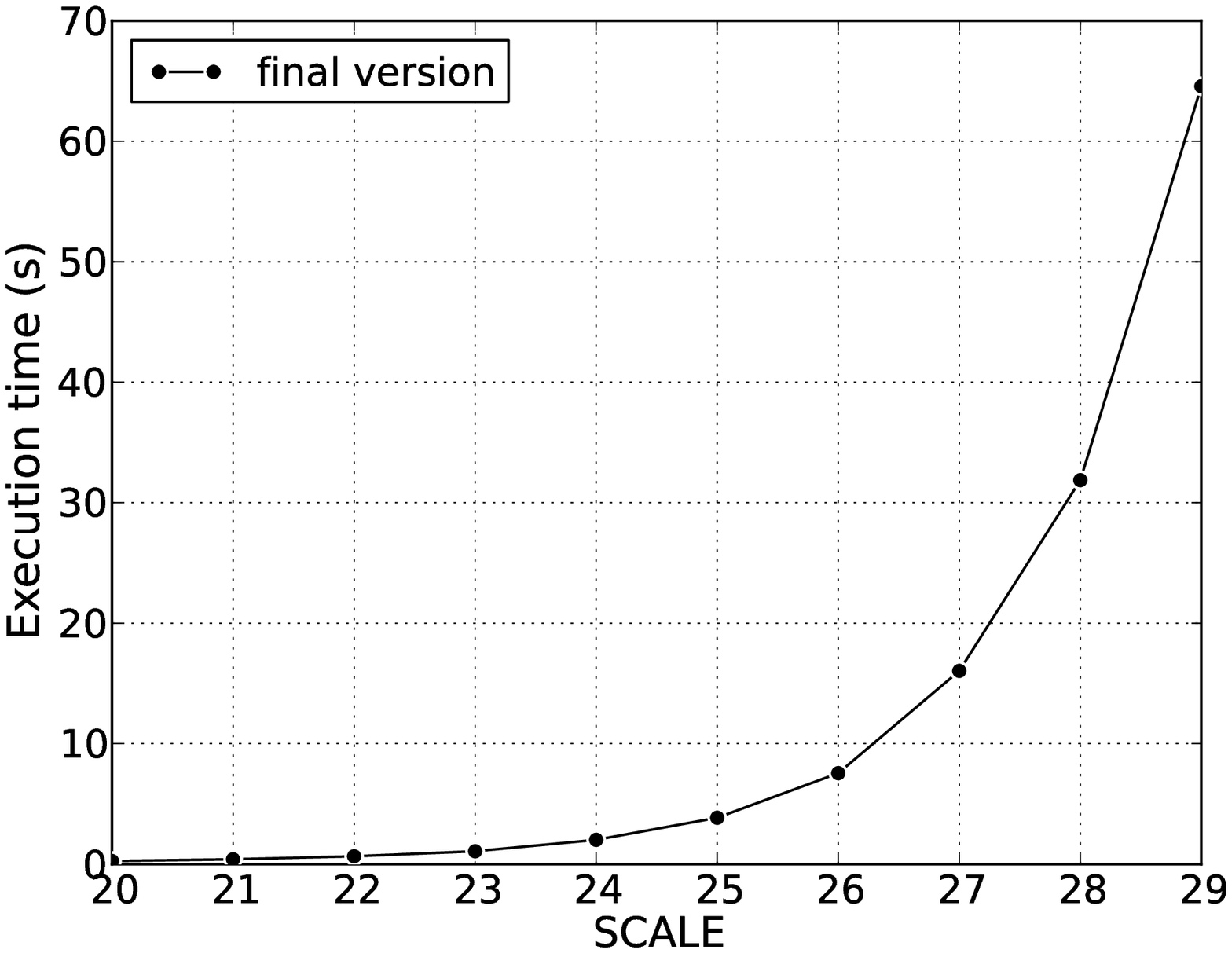}
\caption{Execution time of the final algorithm version for graphs of different sizes. 32 nodes of the MVS-10P cluster. 8 MPI processes per node.}
\label{ris:best_v_time_diff_graphs}
\end{minipage}
\end{center}
\end{figure}

In~Fig.~\ref{ris:avg_msg_size} we show the dependence between average size of communication messages and execution time of the final algorithm version. Here the message size refers to an aggregated message sent over the interconnect. The value of the $MAX\_MSG\_SIZE$ aggregation parameter is 20000 bytes. The figure shows that with increasing number of nodes the message size decreases. On 32 nodes messages are short; their size does not exceed 2 KB. It is also clear that the size of messages depends on the algorithm execution time. 


We suppose that the main limitation factor of the algorithm performance can be latency or injection rate of short messages.    

Fig.~\ref{ris:best_v_time_diff_graphs} shows the weak scaling for RMAT graphs on 32 nodes of the MVS-10P cluster. RMAT-29 is the largest graph that fits into the memory of 32 nodes; it takes a total of 205 GB. It should be noted that the implementation of the algorithm for solving an MST problem is scalable in-memory, i.e. with an increase in the number of nodes it is possible to increase the size of the graph.

\section{Conclusion}
The paper presents the parallel algorithm for finding minimum spanning tree (forest) in the graph for distributed memory systems, and the algorithm implementation that has been made using MPI. 

Compared with the original GHS algorithm the proposed parallel algorithm has the following key features:
\begin{itemize}
\item the requirement of message processing order has been relaxed for \textit{Test} messages, which doubled the scaling of the algorithm;
\item algorithm is generalized for the case of processing a disconnected graph and builds a minimum spanning forest, while the original algorithm is only applicable to connected graphs.
\end{itemize}

As well the presented algorithm adopts some optimization techniques, namely hashing as the local edge search and the compression of communication messages. The algorithm implementation linearly scales on 32 nodes of the MVS-10P Infiniband cluster.

In the next paper edition we plan to present extended performance evaluation of the proposed algorithm and to study the main limiting factors of the algorithm using LogGOPS model and large-scale applications simulator. In the future we plan to improve algorithm scaling and develop hybrid MPI+OpenMP implementation of the algorithm.

\end{document}